\documentclass[fleqn,12pt,twoside]{article}
\usepackage{espcrc1}

\usepackage{graphicx}
\usepackage{epsfig}
\usepackage[figuresright]{rotating}

\newcommand{\AmS}{{\protect\the\textfont2
  A\kern-.1667em\lower.5ex\hbox{M}\kern-.125ens}}

\title{Abundance Ratios in the Galactic Bulge and Super Metal-Rich Type II Nucleosynthesis:  Pitfalls of the Analysis
\thanks{Based on data obtained at the W. M. Keck Observatory, which is operated
as a scientific partnership among the California Institute of Technology, 
the University of California, and NASA, and was made possible by the generous
financial support of the W. M. Keck Foundation.}}

\author{Jon P. Fulbright\address[ciw]{Observatories of the Carnegie Institution of Washington\\Pasadena, California, USA}
R. Michael Rich\address[ucla]{University of California, Los Angeles\\ Los Angeles, California, USA}
Andrew McWilliam\addressmark[ciw]}
       
\begin{document}

% typeset front matter
\maketitle

\begin{abstract}
We present abundance results from our Keck/HIRES observations of 
giants in the Galactic Bulge.  We confirm that the metallicity distribution
of giants in the low-reddening bulge field Baade's Window can be well-fit
by a closed-box enrichment model.  We also confirm previous observations 
that find enhanced [Mg/Fe], [Si/Fe] and [Ca/Fe] for all bulge giants, 
including those at super-solar metallicities.  However, we find that the
[O/Fe] ratios of metal-rich bulge dwarfs decrease with increasing metallicity,
contrary to what is expected if the enhancements of the other $\alpha$-elements
is due to Type II supernovae enrichment.  We suggest that the decrease in 
oxygen production may be due to mass loss in the pre-supernova evolution of
metal-rich progenitors.
\end{abstract}

\section{INTRODUCTION}

The Galactic Bulge contains about 20 percent of the Galaxy's
stellar mass.  Theories of its formation include a primoridal
free-fall collapse, remnants of accretion espisodes, or secular
evolution of bar instabilities (\cite{f03}).  Accurate stellar
abundance determinations can help distingush between these models.

Early investigations of bulge stars \cite{wr83,r88} found a range of 
metallicity in the bulge, including a population of metal-rich stars.
McWilliam \& Rich \cite{mr94} obtained the moderate-resolution echelle
data for 11 bulge K giants.  The analysis of these spectra found 
enhancements in the [X/Fe] ratios for the $\alpha$-elements Mg and Ti,
including stars with overall metallicities greater than solar.  These
enhancements can be best explained by a rapid chemical evolution history 
dominated by Type II supernovae \cite{m99}.  

The limited quality and sample size of the McWilliam \& Rich data demanded
a renewed effort to study stellar abundances in the bulge.  
We have been engaged in a long-term program to study the composition
of stars in the Galactic bulge, with the aim of constraining the conditions
of the bulge's formation and chemical evolution at a level of detail that
is impossible to obtain with any other method.  

%While only data from one
%bulge field is presented here, we have obtained data in two other fields
%of different Galactic latitude.  This will enable the homogeneity of the

\section{DATA}

The observational data consist of Keck/HIRES spectra of 28 bulge 
giants observed between 1998 and 2001.  All of the target stars are
located within Baade's Window, a region of low reddening located 
about 4 degrees from the Galactic center.
%(the line-of-sight passes
%within $\sim 500$ pc of ).  

The spectra have a resolving power of 45,000 to 60,000. 
Typical signal-to-noise levels are
about 50:1.  The spectrograph settings allow for a wavelength
coverage of approximately 5400--7800 \AA, but there are considerable gaps in the
coverage toward the red.  The spectra were reduced using T. Barlow's
pipeline reduction program MAKEE.

In addition to the bulge stars several bright local disk giants were
observed using the 2.5-m du Pont telescope and its echelle spectrograph.
Although
the resolution of these spectra is slightly lower than the Keck data, the 
signal-to-noise ratio of these spectra generally exceed 200:1.  
These giants were picked to have similar stellar parameters
as the bulge giants in order to act as a comparison sample.  
%The spectra
%of these stars were analyzed in the same manner as the bulge stars.

\section{ANALYSIS}

The abundance analysis employs the LTE 1-D spectrum synthesis program
MOOG \cite{sn73} and the grid of LTE model atmospheres from the Kurucz
web page (http://kurucz.harvard.edu).  The Fe line list was created by
searching the Kurucz line list for all Fe lines in the wavelength region
that are unblended and of the appropriate range of strengths.   Many of these
Fe lines do not have laboratory gf-values available, so we conducted our
analysis differentially from Arcturus, a well-studied nearby giant.  This should
help reduce any systematic errors in the analysis related to atomic data
and stellar atmospheres.

Additional details on the analysis method, including determining continuum
regions, finding stellar parameters, and the details of the abundance 
analysis of the non-iron lines will be presented in a series of papers
in the near future.

\section{RESULTS}

\subsection{Metallicity Distribution Function}

The metallicity distribution function (MDF) is an indicator of
the enrichment history of a system.  Large samples can be obtained
\cite{r88,sdr}
using low resolution spectra in order to measure the MDF of Baade's Window.
Sadler et al. found that the mean metallicity is greater than solar and 
the MDF is well-fit by a closed-box gas exhaustion model.  

Our high-resolution sample can be used to recalibrate the earlier 
low-resolution works.  When this is done to the 268 stars of the Sadler
et al. sample, we find that our recalibration finds an exponential 
distribution of Z/Z$_{\odot}$ similar to that found in the earlier results.
However, this preliminary
recalibration assumes Z goes as [Fe/H], which may not be fully accurate.
As we will see later Baade's Window giants to not show solar ratios
of many of the alpha elements.  Correcting for this would increase the
values of Z for bulge stars.

\subsection{Oxygen and Magnesium Abundances}

Oxygen is the third most common element in the Universe behind only
hydrogen and helium.  Enhanced [O/Fe] ratios have traditionally been
used to indicate a population heavily enriched by Type II supernovae.
If the bulge enrichment pattern was dominated by Type II supernovae,
as indicated by the enhancement of other alpha elements 
(bottom panel of Figure 1), then the [O/Fe] ratios as well 
should be uniformly enhanced in bulge stars.  However, in Figure 1 we see
that the [O/Fe] distribution among bulge stars follows the same trend 
seen in the local disk giants.

\centerline{\psfig{file=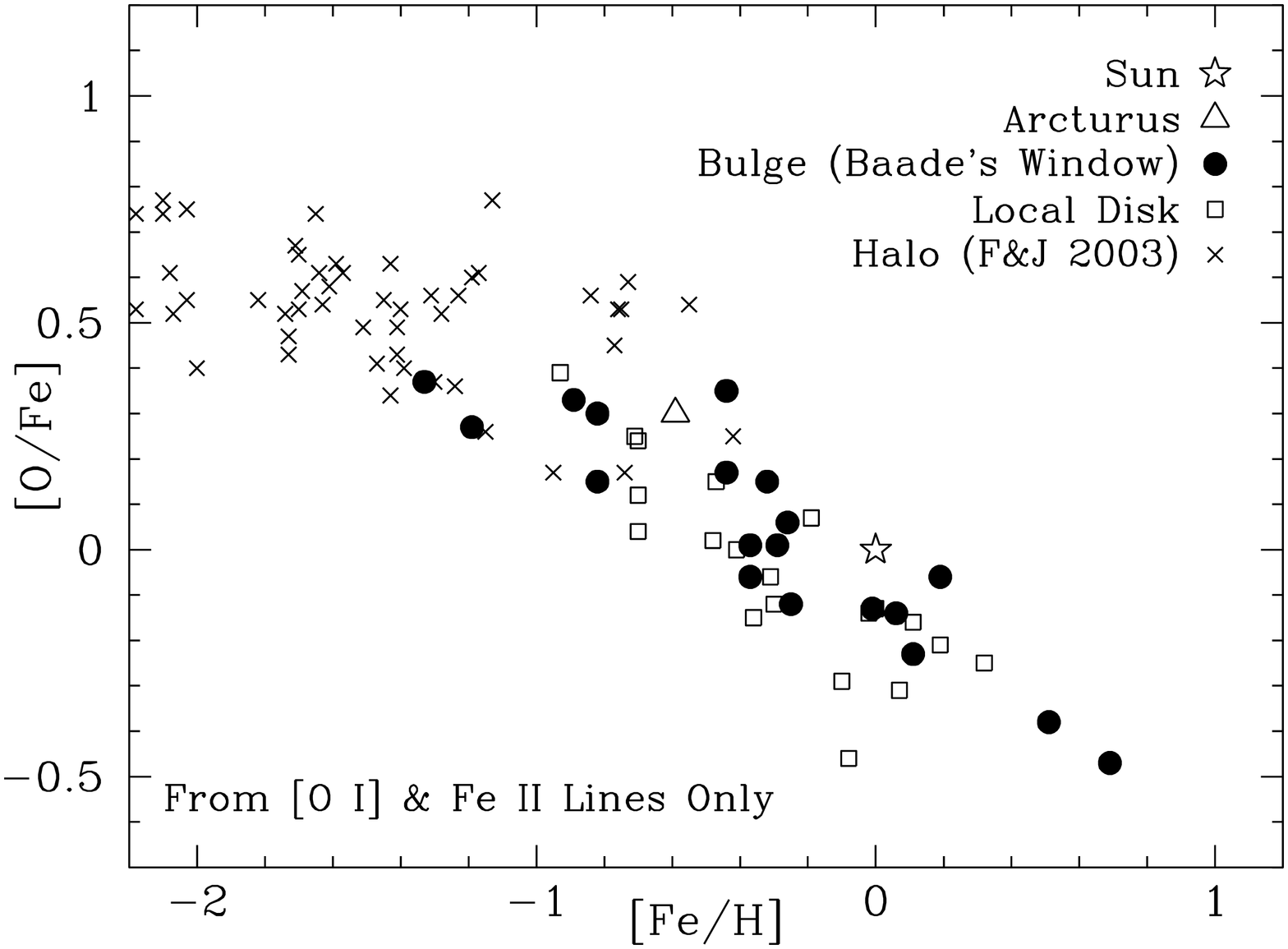,height=4.9in,angle=270}}
\centerline{\psfig{file=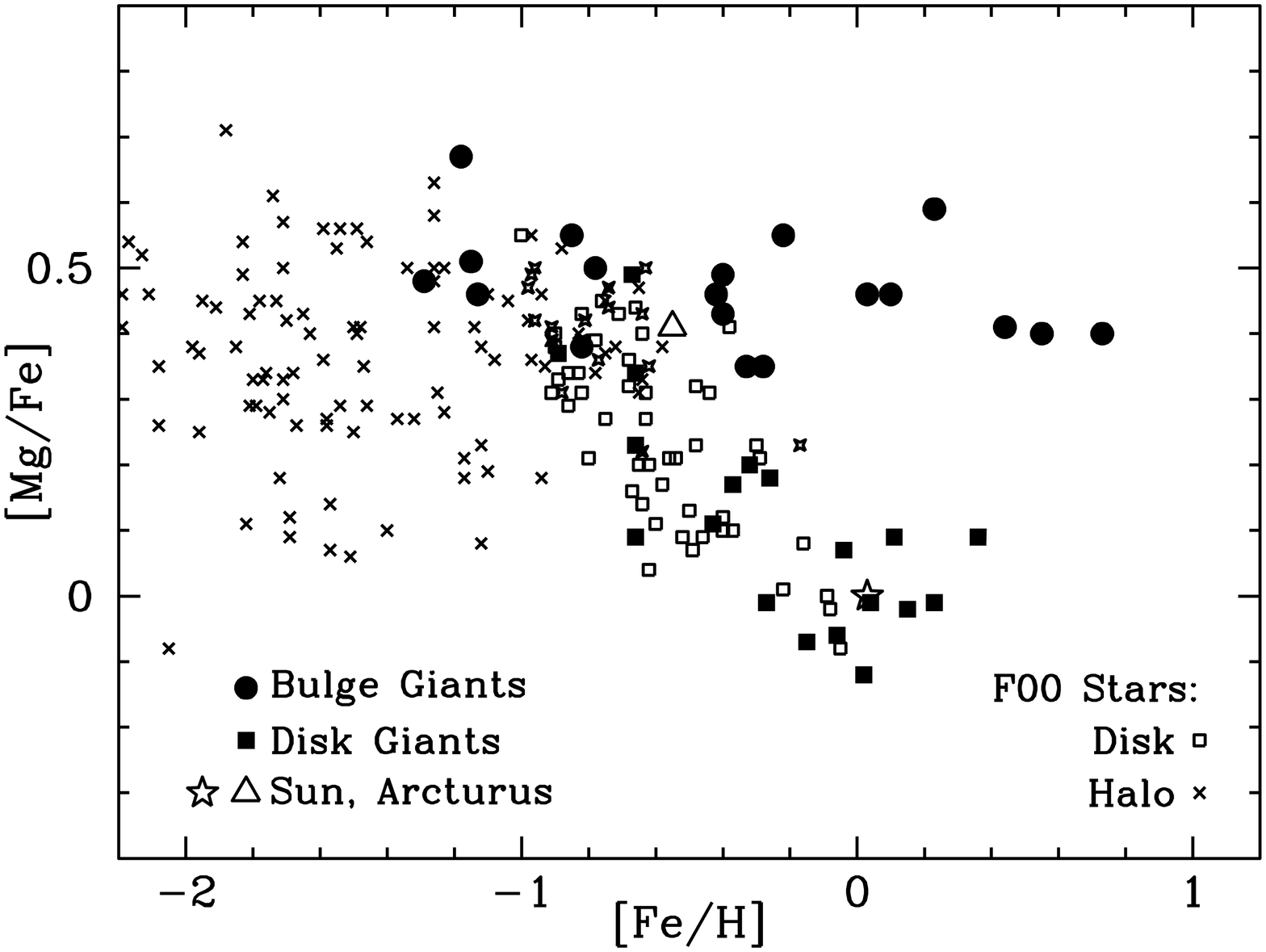,height=4.9in,angle=270}}
\footnotesize 
{\bf Figure 1:} The distribution of [O/Fe] (top) and [Mg/Fe] (bottom) vs. 
[Fe/H] for stars of various populations.  The solid circles denote the 
Baade's Window sample, the solid squares the local disk giant comparison
sample, and the smaller shapes literature samples of various populations.
The literature data come from \cite{f00,fj03}
%In the top diagram the bulge oxygen abunduances parallel the pattern seen
%in solar neighborhood disk and halo stars.  However, in the lower diagram
%the enhanced [Mg/Fe] ratios in metal-rich bulge stars indicate that the
%primary enrichment mechanism in these stars is Type~II supernovae.  
%Type~II model predict that oxygen and magnesium should be produced 
%in similar progenitors.  
%\clearpage
\vspace{0.2in}

%We plot the [O/Fe] ratio as a function of [Fe/H] in Figure 1.  
%Contrary to our predictions, the [O/Fe] ratio drops at higher
%metallicitiy, almost exactly following the trend seen in  disk stars.

\normalsize

The bulge distribution of the other $\alpha$-elements follows the pattern 
predicted to be produced by Type II supernovae.  This creates a quandary:  
Type II models (e.g. \cite{ww95})  predict that O and Mg
are produced in similar mass progenitors.  There are no major producers
of Mg other than Type II supernovae.  Therefore, why do O and Mg
not show similar distributions?

A possible solution may be due to the fact that while O and Mg should
come from similar stars, they are not formed in similar layers within
those stars.  Oxygen
is produced during hydrostatic helium burning while magnesium comes from
hydrostatic carbon and neon burning.  In high-metallicity high-mass
stars, post main-sequence mass loss can remove a large fraction of the
star's original mass.  During this Wolf-Reyet phase it may be possible
that the final hydrostatic He-burning layer could be greatly reduced or
even lost completely while the C-burning layer could be preserved.
The Wolf-Reyet progenitor models of ref. \cite{ww95} do not
predict this amount of mass loss, so further work is necessary to confirm
this hypothesis.

\section{FUTURE DIRECTIONS}

The analysis of the bulge data continues.  The spectra exhibit absorption
lines of many elements not discussed here, including C, N, Na, Al, Fe-group
elements and many elements heavier than the Fe-group, including several of
the traditional indicators of the s- and r-processes.  In addition, we have
obtained data in three other fields at different Galactic latitudes.  This
will allow for the investigation of the homogeneity of the chemical 
properties of the bulge to be studied, which may reveal evidence of
galactic accretion or multiple star formation events.

%\centerline{\psfig{file=ZbinBW.ps,height=5.0in,angle=0}}
%\footnotesize 
%{\bf Figure 1:} The recalibrated metallicity distribution function (MDF) for 
%Baade's Window based on the sample of Sadler et al. (1996).  The is well
%fit by the closed-box gas exhaustion simple model.  The mean abundance 
%over twice the mean abundance of the disk in the solar neighborhood.

%%%%%%%%

%References should be collected at the end of your paper. Do not begin
%them on a new page unless this is absolutely necessary. They should be
%prepared according to the sequential numeric system making sure that
%all material mentioned is generally available to the reader. Use
%\verb+\cite+ to refer to the entries in the bibliography so that your
%accumulated list corresponds to the citations made in the text body. 
%
%Above we have listed some references according to the
%sequential numeric system \cite{Scho70,Mazu84,Dimi75,Eato75}.
\end{document}